\documentclass[useAMS,usenatbib]{mnras}
\usepackage{graphicx,natbib,amssymb,amsmath,stmaryrd,makecell}
\usepackage{txfonts}
\usepackage{xcolor}
\usepackage{hyperref}
\usepackage{xspace,dsfont}
\usepackage{caption, lipsum, comment}

\graphicspath{{Figures/}}

\newcommand{\expf}[1]{{{\rm e}^{#1}}}

\newcommand{\nbb}{{n^{\rm pl}}}

\newcommand{\vgh}{{\hat{\boldsymbol\gamma}}}
\newcommand{\vghp}{{\hat{\boldsymbol\gamma}'}}

\newcommand{\vb}{{\boldsymbol{\beta}}}
\newcommand{\vbh}{{\boldsymbol{\hat{\beta}}}}

\newcommand{\gammac}{{\gamma_{\rm p}}}

\newcommand{\id}{{\,\rm d}}

\newcommand{\beq}{\begin{equation}}   %

\newcommand{\eeq}{\end{equation}}   %

\newcommand{\beqa}{\begin{eqnarray}}   %

\newcommand{\eeqa}{\end{eqnarray}}   %

\newcommand{\bealf}[1]{\begin{align} #1 \end{align}}

\newcommand{\beal}{\begin{align}}
\newcommand{\enal}{\end{align}}

\newcommand{\bspl}{\begin{split}}

\newcommand{\espl}{\end{split}}

\newcommand{\bsub}{\begin{subequations}}

\newcommand{\esub}{\end{subequations}}

\newcommand{\bmulti}{\begin{multline}}   %

\newcommand{\beqm}{\begin{mathletters}}   %

\newcommand{\eeqm}{\end{mathletters}}   %

\newcommand{\me}{m_{\rm e}}

\newcommand{\Ne}{N_{\rm e}}

\newcommand{\Te}{T_{\rm e}}

\newcommand{\Tg}{T_{\gamma}}

\newcommand{\The}{\theta_{\rm e}}

\newcommand{\sigT}{\sigma_{\rm T}}

\newcommand{\vek} [1]{\mbox{\boldmath${#1}$\unboldmath}}

\newcommand{\sterling}[2]{\genfrac{[}{]}{0pt}{0}{#1}{#2}}
\newcommand{\eulerian}[2]{\genfrac{\langle}{\rangle}{0pt}{0}{#1}{#2}}

\newcommand{\oDnu}{{\mathcal{\hat{D}}_{\nu}}}

\newcommand{\oOnu}{{\mathcal{\hat{O}}_{\nu}}}

\newcommand{\oOx}[1]{{x^{#1}\partial_x^{#1}}}

\renewcommand{\Ne}{N_{\rm e}}

\usepackage{comment}

\title[Boost operator approach to rSZ]
{Boost operator approach to the relativistic SZ effect}

\author[]
{
Jens~Chluba$^{1}$\thanks{E-mail:Jens.Chluba@Manchester.ac.uk}
and
Erik~Rosenberg$^{1}$
\\
$^{1}$Jodrell Bank Centre for Astrophysics, Alan Turing Building, University of Manchester, Manchester M13 9PL
}

\begin{document}

\date{\vspace{-3mm}
{Accepted 2025 --. Received 2025}}

\maketitle

\begin{abstract}
The Sunyaev-Zeldovich (SZ) effect provides a powerful cosmological probe. We demonstrate that the corresponding relativistic SZ signal can be accurately calculated using the recently developed boost operator approach. We obtain formally exact expressions for the required differential operator that can be used to generate relativistic temperature and velocity correction functions to any order. 
Many of the otherwise cumbersome intermediate steps can be avoided and the required boost operator elements can be quickly generated using recurrence relations of the underlying aberration kernel. 
We confirm previous analytic expressions describing the relativistic SZ effect and give new expressions at third order in the cluster's peculiar velocity, demonstrating the feasibility of the boost operator method.
Our derivation also highlights general properties of the boost operator and showcases its application to radiative transfer problems of broader interest.
\end{abstract}


\section{Introduction}
\label{sec:Intro}
The Sunyaev-Zeldovich (SZ) effect is long-known to be a powerful cosmological probe \citep{Birkinshaw1999, Carlstrom2002, Mroczkowski2019}. Rapid progress is being made on the observational frontier \citep[e.g.,][]{Planck2013SZ, Bleem2015, PlanckSZ2016, ACTSZ2025} and upcoming CMB experiments \citep{SOWP2018, SimonsII2025} promise the detections of tens of thousands of clusters through the SZ effect.

While the leading order thermal and kinematic SZ effect signals predictions have been made long ago \citep{Zeldovich1969, Sunyaev1980}, more accurate computations that account for relativistic temperature and velocity corrections have only been studied later
\citep[e.g.,][]{Wright1979, Rephaeli1995, Pointecouteau1998,
Itoh98, Sazonov1998, Challinor1998, Nozawa1998SZ, Challinor1999, Sazonov1999, Shimon2004,
Chluba2005b}. With modern computational methods, highly accurate representations of the SZ signal can be reliably obtained using {\tt SZpack} \citep{Chluba2012SZpack, Chluba2012moments, Lee2024SZpack}, essentially providing an exact extendable framework for SZ modeling.

One of the approximate methods for SZ modeling is the so-called asymptotic expansion in orders of the electron temperature $\The=k\Te/\me c^2$ and cluster peculiar velocity, $\beta_{\rm p}=\varv_{\rm p}/c$ \citep[e.g.,][]{Nozawa2005, Chluba2012SZpack}. The related expressions look complicated and even with algebra packages such as {\tt Mathematica}, the related derivations are quite time-consuming at high order, requiring  integrations of long lists of terms appearing in the Taylor expansions of the scattering cross section. 

In this paper we investigate a novel derivation of the relativistic SZ effect using the boost operator approach recently studied by \citep{ChlubaBO25}. The boost operator (composed of frequency derivatives $\oOnu=-\nu \partial_\nu$) provides an exact analytic framework for computing the Lorentz transformations of photon signals to all order in velocity, as also considered by \citet{Yasini2017} in less general form. The boost operator is based on the simpler aberration kernel \citep{Challinor2002, Chluba2011ab, Dai2014}, which describes how spherical harmonic coefficients transform between moving frames and can be computed efficiently using exact recurrence relations.

As we show here, the boost operator opens the possibility to obtain an {\it exact SZ operator} to all orders of the electron temperature and cluster's peculiar velocity (see Sect.~\ref{sec:exact_all}). The final computation is quite simple since recurrence relations with standard Taylor series expansions can be used to obtain the result term by term. We confirm existing analytic expressions and also give some of the expressions to third order of the cluster motion. Even if somewhat of academic interest, our derivation clarifies the origin of various terms and coefficients as well as showcases the unique properties and applications of the boost operator to radiative transfer problems.

The paper is structured as follows: In Sect.~\ref{sec:rSZ}, we treat the thermal SZ effect to all orders in the temperature. In Sect.~\ref{sec:exact_all}, we then generalize to include all kinematic corrections. Our derivations rely heavily on the expressions given in \citet{ChlubaBO25} and \citet{Dai2014}, and we recommend consulting those papers for details. A brief summary of our findings is given in Sect.~\ref{sec:Conc}.

\vspace{-3mm}
\section{Exact treatment of the thermal SZ effect}
\label{sec:rSZ}
To obtain the thermal SZ effect including all relativistic temperature corrections, we can think about the scattering problem as a superposition of individual boosts in separate directions. For the scattering of CMB photons by hot thermal electrons, the  interaction in the electron rest frame can be computed using the Thomson collision term, which omits Klein-Nishina corrections and recoil (see Appendix~\ref{app:Thomson}). This is possible because CMB photons have energies $h\nu/\me c^2\ll 1$ even in the rest frame of the moving electron.

For the scattering of photons by an electron moving in the direction $\vbh$ at speed $\beta =\varv/c$ using the boost operator approach, we then have \citep{ChlubaBO25}:
\bealf{
\label{eq:dn_transformed}
\left.\frac{\id  n(\nu,\vgh, \vb)}{ \id \tau}\right|_{\rm T}
&=\frac{1}{\gamma}\sum_{\ell,\ell'} Y_{\ell0}(\vgh)\,{^{-1}}\mathcal{\hat{B}}^0_{\ell 0}(\nu, -\beta)\,{^0}\mathcal{\hat{B}}^0_{0\ell'}(\nu, \beta) \,n_{\ell' 0}(\nu)
\nonumber\\
&\!\!\!\!\!\!\!\!\!\!\!\! + \frac{1}{\gamma}\sum_{\ell, m, \ell'} \frac{Y_{\ell m}(\vgh)}{10}
\,{^{-1}}\mathcal{\hat{B}}^m_{\ell 2}(\nu, -\beta)\,{^0}\mathcal{\hat{B}}^m_{2\ell'}(\nu, \beta) \,n_{\ell' m}(\nu)
\\
\nonumber
&\!\!\!\!\!\!\!\! - \sum_{\ell, m}Y_{\ell m}(\vgh)
\,\Big\{n_{\ell m}(\nu)-\beta
\left[C^m_{\ell+1}\,n_{\ell+1, m}(\nu)+C^m_{\ell} \,\,n_{\ell-1, m}(\nu)\right]
\Big\}
}
in the lab frame, where all sums over $m$ are determined by the minimal value of $\ell$ appearing. 
Here, $n(\nu,\vgh, \vb)$ denotes the photon occupation number at observing frequency $\nu$ and for a photon propagating in the direction $\vgh$; $Y_{\ell m}$ gives the spherical harmonic functions; $\tau$ is the Thomson optical depth along the line of sight; $\gamma=1/\sqrt{1-\beta^2}$ is the Lorentz factor of the electron, and ${^{d}}\mathcal{\hat{B}}^m_{\ell \ell'}$ is the boost operator for Doppler-weight $d$ and spin-weight $s=0$, coupling multipoles $\ell$ and $\ell'$ at fixed $m$ (for $\vbh\parallel z$-axis). We also have the coefficients $C^m_{\ell}=\sqrt{(l^2-m^2)/(4\ell^2-1)}$, appearing in recurrence relations for the spherical harmonic functions.

For the SZ effect, we assume that in the lab frame the unscattered photon occupation number is isotropic and Planckian, i.e., $$n(\nu,\vgh, \vb)=Y_{00}\,n_{00}(\nu)=\frac{1}{\expf{h\nu/k \Tg}-1}\equiv \nbb(\nu).$$ 
This means that we can consider the simpler problem\footnote{This expression is valid in the single scattering limit as anisotropies in the photon fluid are generated after one scattering.}
\bealf{
\label{eq:dn_transformed_SZ}
\left.\frac{\id  n(\nu,\vgh, \vb)}{ \id \tau}\right|_{\rm T}
&=\Bigg\{\sum_{\ell} Y_{\ell0}(\vgh)\,\left[\mathcal{\hat{D}}_{\ell 0}
+
\frac{\mathcal{\hat{D}}_{\ell 2}}{10}-\delta_{\ell 0}+\frac{\beta}{\sqrt{3}}\,\delta_{\ell 1}
\right]\Bigg\}\,n_{00}(\nu)
\nonumber \\
\mathcal{\hat{D}}_{\ell \ell'}(\nu, \beta)&=
{^{-1}}\mathcal{\hat{B}}^0_{\ell \ell'}(\nu, -\beta)\,{^0}\mathcal{\hat{B}}^0_{\ell'0}(\nu, \beta)/\gamma,
}
where we introduced the operators $\mathcal{\hat{D}}_{\ell \ell'}$ for convenience. This expression assumes that $\vbh$ is aligned with the $z$-axis. Because the thermal electron distribution is isotropic, however, we can obtain the average over all directions $\id\vbh$ by  averaging over all $\id\vgh$:\footnote{This is more formally shown below with Eq.~\eqref{eq:av_Y_gamma}.}
\bealf{
\label{eq:dn_transformed_SZ_single_p}
\left.\frac{\id  n(\nu,\beta)}{ \id \tau}\right|_{\rm T}
&=\int \frac{\id \vgh}{4\pi}\left.\frac{\id  n(\nu,\vgh, \vb)}{ \id \tau}\right|_{\rm T}
\nonumber\\
&=
\Bigg\{\mathcal{\hat{D}}_{00}(\nu, \beta)
+
\frac{\mathcal{\hat{D}}_{02}(\nu, \beta)}{10}
-1\Bigg\}\,\nbb(\nu)
}
leaving only the monopole terms. Defining the SZ operator
\bealf{
\label{eq:S_single}
\mathcal{\hat{S}}_{\rm th}(\nu, p)&=\mathcal{\hat{D}}_{00}(\nu, p/\gamma)
+
\frac{\mathcal{\hat{D}}_{02}(\nu, p/\gamma)}{10}
-1
}
with electron momentum $p=\beta \gamma$ and Lorentz factor $\gamma = \sqrt{1+p^2}$, we next have to average over the electron speed to obtain the final exact result. For the electron momentum distribution we use the relativistic Maxwellian:
\begin{equation}
f(\gamma) = \frac{\exp\Big(-\gamma/\The\Big)}{\theta_{\rm e} K_2(1/\theta_{\rm e})}
\quad \text{with} \quad \theta_{\rm e} = \frac{kT_{\rm e}}{m_{\rm e}c^2}.
\label{eq:rmbdist}
\end{equation}
Here $K_2(1/\theta_{\rm e})$ is the modified Bessel function of second order, necessary to ensure the normalization $\int_0^{\infty}p^2f(\gamma)\,\text{d}p = 1$.
This means that we can compute the relativistic thermal SZ effect using the thermal SZ operator
\bsub
\bealf{
\label{eq:dn_transformed_SZ_single_av}
\Delta n\big|_{\rm th}
&\approx \tau \,\mathcal{\hat{S}}_{\rm th}(\nu, \The)\,\nbb(\nu)
\\
\mathcal{\hat{S}}_{\rm th}(\nu, \The)&=
\int_0^\infty
p^2 f(\gamma)\,\text{d}p\,
\Bigg\{\mathcal{\hat{D}}_{00}(\nu, \beta)
+
\frac{\mathcal{\hat{D}}_{02}(\nu, \beta)}{10}
-1\Bigg\}.
}
\esub
Formally, this is the exact solution to the relativistic thermal SZ effect at {\it all} orders in $\The$.

In practice, the thermal average can be carried out after performing a Taylor series of the single momentum operator in orders of $p$, which can be easily done with {\tt Mathematica}. For this we need expressions for $\mathcal{\hat{D}}_{00}(\nu, \beta)$ and $\mathcal{\hat{D}}_{02}(\nu, \beta)$. Given the boost operator identity \citep{ChlubaBO25} 
\bealf{
{^d}\mathcal{\hat{B}}^m_{\ell \ell'}(\nu, \beta)\equiv {^{d+\oOnu}}\mathcal{K}^m_{\ell \ell'}(-\beta)
}
with $\oOnu=-\nu\partial_\nu$, this means we only need the aberration kernel elements \citep{Dai2014, ChlubaBO25}
\bsub\bealf{
\label{eq:Kernels_0_2}
{}^{d}\mathcal{K}_{00}^{0}(-\beta)
&=\frac{(\gamma+p)^{1-d}-(\gamma-p)^{1-d}}{2(1-d)p}
\\
^{d}\mathcal{K}_{20}^{0}(-\beta)
&=\frac{3\sqrt{5}}{2p^2}
\left[
\left(1+\frac{2}{3}p^2\right)\,^{d}\mathcal{K}_{00}^{0}
-2\gamma\,^{d-1}\mathcal{K}_{00}^{0}+
{}^{d-2}\mathcal{K}_{00}^{0}\right]
\\[1mm]
^{d}\mathcal{K}_{02}^{0}(-\beta)
&\equiv{}^{2-d}\mathcal{K}_{20}^{0}(\beta),
}
\esub
which we computed as explained in Appendix~\ref{app:Kernel_raise}.
We then find
\bsub
\bealf{
\label{eq:D_ops}
\mathcal{\hat{D}}_{00}(\nu, \beta)&=
^{\oOnu-1}\mathcal{K}^0_{00}(\beta)\,{}^{\oOnu}\mathcal{K}^0_{00}(-\beta)/\gamma
\nonumber\\
&\equiv
\frac{(\gamma+p)^{3-2\oOnu}+(\gamma-p)^{3-2\oOnu}-2\gamma}{4(2-\oOnu)(1-\oOnu) \, \gamma p^2}
\\
\mathcal{\hat{D}}_{02}(\nu, \beta)&=
\frac{^{\oOnu-1}\mathcal{K}^0_{02}(\beta)\,{}^{\oOnu}\mathcal{K}^0_{20}(-\beta)}{\gamma}\equiv
\frac{^{3-\oOnu}\mathcal{K}^0_{20}(-\beta)\,{}^{\oOnu}\mathcal{K}^0_{20}(-\beta)}{\gamma}
}
\esub
omitting the explicit expression for $\mathcal{\hat{D}}_{02}(\nu, \beta)$.\footnote{Here, we used the identity $\gamma - p \equiv (\gamma +p)^{-1}$ to simplify the expressions. In the massaging of terms, it is furthermore useful to introduce $p_\pm=\gamma\pm p$ and then replace $p=\frac{1}{2}(p_+-p_-)$ and $
\gamma=\frac{1}{2}(p_++p_-)$ to group terms, a procedure that was also followed by \citet{CSpack2019}.}
Because the required operators are simple functions in $p$ it is easy to generate the single-momentum SZ operator in terms of $\oOnu$.
As an example, we find
\bealf{
\label{eq:S_single_4}
\mathcal{\hat{S}}_{\rm th}(\nu, p)&\approx \left[\oOnu^2-3\oOnu\right]\frac{p^2}{3}+
\left[
\frac{42}{5}\oOnu+\frac{7\oOnu^2}{2}-\frac{21\oOnu^3}{5}+\frac{7\oOnu^4}{10}\right] \frac{p^4}{15}
\nonumber\\
&\!\!\!\!\!\!\!\!\!\!\!\!\!\!\!\!\!\!
-
\left[
44\oOnu+\frac{473\oOnu^2}{15}-\frac{209\oOnu^3}{10}-\frac{143\oOnu^4}{30}+\frac{33\oOnu^5}{10}-\frac{11\oOnu^6}{30}\right]\frac{p^6}{105}
\nonumber \\
&\!\!\!\!\!\!\!\!\!\!\!\!\!\!
+\Bigg[
\frac{2304\oOnu}{7}+\frac{10176\oOnu^2}{35}-\frac{4736\oOnu^3}{35}-\frac{7856\oOnu^4}{105}+\frac{192\oOnu^5}{7}
\nonumber \\
&\!\!\!\!\!\!\!\!\!\!
+\frac{352\oOnu^6}{105}-\frac{64\oOnu^7}{35}+\frac{16\oOnu^8}{105}\Bigg]\frac{p^8}{945}
}
up to $8^{\rm th}$ order in $p$. Higher orders can be obtained in a similar manner. The next step is then to evaluate the moments
\begin{align}
\label{app:MMrMB_def}
\left<p^k\right>&=\int p^{k+2} f(\gamma) \id p = \int \gamma \sqrt{\gamma^2-1}^{k+1} f(\gamma) \id \gamma
\end{align}
which are given by \citep[see][]{CSpack2019} 
\begin{align}
\label{app:MMrMB_k}
\left<p^{k}\right>&=\frac{2\,(2\The)^{k/2} K_{(k+4)/2}(1/\The)}{\sqrt{\pi}K_2(1/\The)}\,\Gamma\left(\frac{k+3}{2}\right),
\end{align}
in terms of modified the Bessel functions, $K_n(x)$, and the $\Gamma$ function. These can again be expanded in orders of $\The$. For example, we have $\left<p^2\right>\approx 3\The \left(1+\frac{5}{2}\The\right)$ and $\left<p^4\right>\approx 15\The^2$ \citep[more examples can be found in Appendix~C of][]{CSpack2019}. 

To obtain the classical expression for the non-relativistic SZ effect \citep{Zeldovich1969}, we note that the lowest order photon diffusion operator, $\oDnu=\nu^{-2} \partial_\nu \nu^4 \partial_\nu$, can be written as $\oDnu=\oOnu^2-3\oOnu$ \citep[e.g.,][]{chluba_spectro-spatial_2023-I}.
We thus simply evaluate $\mathcal{\hat{S}}_{\rm th}$ to first order in $\The$ and make this substitution to directly find:
\bealf{
\label{eq:dn_transformed_SZ_single_av}
\mathcal{\hat{S}}_{\rm th}(\nu, \The)&
\approx \The \oDnu\quad
\rightarrow \quad
\Delta n\big|_{\rm th}
\approx \tau \The Y_0(x)
}
with $Y_0(x)=\oDnu \nbb(\nu)=\frac{x \expf{x}}{(\expf{x}-1)^2}\left[x \coth(x/2)-4\right]$ and $x=h\nu/k\Tg$.
To derive higher order temperature terms, we aim at obtaining the related distortion spectra 
$Y_n(x)$ \citep[e.g.,][]{Itoh98, Chluba2012SZpack}, such that the relativistic thermal SZ signal can be written as
\bealf{
\label{eq:dn_Itoh}
\Delta n\big|_{\rm th}
\approx \tau \The \sum_k \The^k Y_k(x).
}
Up to $10^{\rm th}$ order in $\The$, the $Y_k(x)$ were computed in \citet{Chluba2012SZpack} and can be given in terms of the derivatives \citep[see Appendix~A2 of][]{Chluba2012SZpack}
\bealf{
\label{eq:xkdkn}
x^k\partial_x^k \nbb&=\frac{(-x)^k\expf{-x}}{(1-\expf{-x})^{k+1}}\sum_{m=0}^{k-1}
\eulerian{k}{m}
\,\expf{-mx}}
of a Planckian. Here, the bracket denotes the Eulerian numbers. To bring the expression for $\mathcal{\hat{S}}_{\rm th}(\nu, p)$ into the required form, we use the operator identity \citep{chluba_spectro-spatial_2023-I} 
\begin{equation}
    \oOnu^k = (-1)^k\sum_{m=1}^k \sterling{k}{m} \,x^m\partial_x^m,
\end{equation}
where the bracket gives the Sterling numbers of second kind. Reorganizing terms in this way, we find
\bealf{
\label{eq:S_single_reorg}
\mathcal{\hat{S}}_{\rm th}(\nu, p)&\approx \left[4\oOx{}+\oOx{2}\right]\frac{p^2}{3}+
\left[
21\oOx{2}+\frac{42\oOx{3}}{5}+\frac{7\oOx{4}}{10}\right] \frac{p^4}{15}
\nonumber\\
&\!\!\!\!\!\!\!\!\!\!\!\!\!\!\!\!\!\!
+
\left[
\frac{616\oOx{3}}{5}+\frac{308\oOx{4}}{5}+\frac{44\oOx{5}}{5}+\frac{11\oOx{6}}{30}\right]\frac{p^6}{105}
\\ \nonumber
&\!\!\!\!\!\!\!\!\!\!\!\!\!\!
+
\left[
768\oOx{4}+\frac{3072\oOx{5}}{7}+\frac{576\oOx{6}}{7}+\frac{128\oOx{7}}{21}+\frac{16\oOx{8}}{105}\right]\frac{p^8}{945}.
}
Collecting all terms at second order in $\The$, we then have
\bealf{
\label{eq:Y1}
Y_1(x)&=\left\{\left[4\oOx{}+\oOx{2}\right]\,\frac{5}{2}
+
\left[
21\oOx{2}+\frac{42\oOx{3}}{5}+\frac{7\oOx{4}}{10}\right]\right\}
\nbb(x)
\nonumber\\
&=\left\{10\oOx{}+\frac{47}{2}\oOx{2}+\frac{42}{5}\oOx{3}
+\frac{7}{10}\oOx{4}\right\}
\nbb(x),
}
confirming the well-known result \citep{Itoh98, Sazonov1998, Challinor1998}. With the same procedure, we confirmed all $Y_k$ given in Appendix~C of \citet{Chluba2012SZpack}. It is also extremely easy to generate higher order function $Y_k$.\footnote{A {\tt Mathematica} file is available at \url{www.chluba.de/Mathematica}} However, it is well-known that this only improves the convergence of the series very slowly \citep[e.g.,][]{Chluba2012SZpack}. For high precision, we recommend using the {\tt combo} function of {\tt SZpack}\footnote{\url{www.chluba.de/SZpack}} or the electron scattering kernel approach \citep[see][for details]{Lee2024SZpack}.

\section{Adding kinematic corrections}
\label{sec:exact_all}
To derive the kinematic SZ effect with all relativistic corrections included, we start by giving the distribution function for moving electrons \citep[e.g.,][]{Lee2024SZpack}
\bealf{
\label{eq:fMB_moving}
f_{\rm p}(\vek{p}) 
=
\frac{(N_{\rm e}/\gammac)\, \exp\left(-\frac{\gammac\gamma}{\The}\right)}{4\pi \The K_2(1/\The) }
\times \exp\left(\frac{\vek{p}_{\rm p} \cdot \vek{p}}{\The}\right),
}
where the cluster's peculiar motion is in the direction $\vbh_{\rm p}$ with a speed $\beta_{\rm p}$ and momentum $\vek{p}_{\rm p}$.
The second exponential term can be expanded in Legendre polynomials as \citep{Lee2024SZpack}
\begin{equation} \label{eqn:boostMB}
\exp\left(\frac{\vek{p}_{\rm p} \cdot \vek{p}}{\The}\right) 
= 
\sqrt{\frac{\pi\,\The}{2 p_{\rm p} p}}\sum_{\ell=0}^\infty (2\ell+1) \,I_{\ell+\frac{1}{2}}\left(\frac{p_{\rm p} p}{\The}\right) P_\ell(\hat{\vek{p}}_{\rm p} \cdot \hat{\vek{p}}),
\end{equation}
where $I_\ell(x)$ is the modified Bessel function of the first kind. Since\footnote{Since $m$=0, we only need the direction cosine of $\vgh$ with respect to the $z$-axis, which is parallel to $\vbh$.}
\bealf{
\label{eq:Y_gamma}
Y_{\ell0}(\vgh)
&\equiv
\sqrt{\frac{2\ell+1}{4\pi}}\,P_{\ell}(\vbh\cdot\vgh)
=\sqrt{\frac{4\pi}{2\ell+1}}\sum_{m=-\ell}^{\ell} Y_{\ell m}(\vgh)\,Y^*_{\ell m}(\vbh)
}
we then have
\bealf{
\label{eq:av_Y_gamma}
\int \id \vbh \,Y_{\ell0}(\vgh)\,P_{\ell'}(\hat{\vek{p}}_{\rm p} \cdot \hat{\vek{p}}) 
&\equiv \sqrt{\frac{4\pi}{2\ell+1}}\,P_{\ell}(\vgh\cdot\vbh_{\rm p}).
}
Collecting terms, we can then introduce the exact SZ operator
\bsub
\label{eq:kSZ_operator}
\bealf{
\mathcal{\hat{S}}_{\rm SZ}(\nu, \The, \vgh, \vb_{\rm p})&=
\sum_{\ell=0}^\infty 
\mathcal{\hat{S}}_{\ell}(\nu, \The, \beta_{\rm p})
\,P_{\ell}(\vgh\cdot\vbh_{\rm p})
\\
\mathcal{\hat{S}}_{\ell}(\nu, \The, \beta_{\rm p})&=
\int_0^\infty
p^2
f_{\ell}(\gamma, \gamma_{\rm p})
\,\mathcal{\hat{S}}_{\ell}(\nu, p)\,\text{d}p
\\
\nonumber 
\mathcal{\hat{S}}_{\ell}(\nu, p)&=
\sqrt{2\ell+1}\left[\mathcal{\hat{D}}_{\ell 0}(\nu, \beta)
+
\frac{\mathcal{\hat{D}}_{\ell 2}(\nu, \beta)}{10}\right]-\delta_{\ell 0}-\beta\,\delta_{\ell 1}
\\
f_{\ell}(\gamma, \gamma_{\rm p}) 
&=
\frac{(N_{\rm e}/\gammac)\, \exp\left(-\frac{\gammac\gamma}{\The}\right)}{\The K_2(1/\The) }\,\sqrt{\frac{\pi\,\The}{2 p_{\rm p} p}}\,I_{\ell+\frac{1}{2}}\left(\frac{p_{\rm p} p}{\The}\right)
}
\esub
which then gives the final SZ signal as
\bealf{
\label{eq:SZ_exact_all}
\Delta n\big|_{\rm SZ}
&\approx \tau^* \mathcal{\hat{S}}_{\rm SZ}(\nu, \The, \vgh, \vb_{\rm p})\,\nbb(\nu)
}
to all orders in $\The$ and $\beta_{\rm p}$. For $\beta_{\rm p}=0$, this reduces to Eq.~\eqref{eq:dn_transformed_SZ_single_av}, as expected.
We note that $\tau^*\equiv \tau/[1-\vb_{\rm p}\cdot \vgh]$ is the optical depth with kinematic corrections included, as we will explain below.

To compute the averaged operators $\mathcal{\hat{S}}_{\ell}(\nu, \The, \beta_{\rm p})$, we shall consider the electron distribution function up to third order in $\beta_{\rm p}$. This yields
\bsub
\bealf{
\label{eq:f_ell_approx}
f_{0}(\gamma, \gamma_{\rm p}) 
&\approx f(\gamma)\left[1-\frac{\beta_{\rm c}^2}{2}+\frac{p^2-3\The \gamma}{6\The^2}\,\beta_{\rm c}^2\right]
\\
f_{1}(\gamma, \gamma_{\rm p}) 
&\approx f(\gamma)\left[1+\frac{p^2-5\The \gamma}{10\The^2}\,\beta_{\rm c}^2\right] \frac{p \beta_{\rm c}}{3\The}
\\
f_{2}(\gamma, \gamma_{\rm p}) 
&\approx f(\gamma)\,\frac{p^2 \beta^2_{\rm c}}{15\The^2}, \qquad
f_{3}(\gamma, \gamma_{\rm p}) 
\approx f(\gamma)\,\frac{p^3 \beta^3_{\rm c}}{105\The^3},
}
\esub
with $f_\ell=0$ for $\ell>3$ at $\mathcal{O}(\beta_{\rm p}^3)$. 
To obtain the operators $\mathcal{\hat{D}}_{\ell 0}(\nu, \beta)$ and $\mathcal{\hat{D}}_{\ell 2}(\nu, \beta)$ for $\ell \leq 3$ we need the aberration kernel elements
\bsub\bealf{
\label{eq:Kernels_kin}
{}^{d}\mathcal{K}_{10}^{0}(-\beta)
&=-\frac{\sqrt{3}}{p}
\left[\gamma\,{}^{d}\mathcal{K}_{00}^{0}-\,{}^{d-1}\mathcal{K}_{00}^{0}\right]
\\
{}^{d}\mathcal{K}_{21}^{0}(-\beta)
&=\frac{\sqrt{3}}{p}
\left[\gamma\,{}^{d}\mathcal{K}_{20}^{0}-\,{}^{d+1}\mathcal{K}_{20}^{0}\right], \,{}^{d}\mathcal{K}_{12}^{0}(-\beta)
\equiv{}^{2-d}\mathcal{K}_{21}^{0}(\beta)
\\
{}^{d}\mathcal{K}_{22}^{0}(-\beta)
&=
\frac{3\sqrt{5}}{2p^2}
\left[
\left(1+\frac{2}{3}p^2\right)\,^{d}\mathcal{K}_{20}^{0}
-2\gamma\,^{d+1}\mathcal{K}_{20}^{0}+
{}^{d+2}\mathcal{K}_{20}^{0}\right]
\\
{}^{d}\mathcal{K}_{30}^{0}(-\beta)
&=-\frac{5\sqrt{7}}{3 p}
\left[\frac{\gamma\,{}^{d}\mathcal{K}_{20}^{0}-\,{}^{d-1}\mathcal{K}_{20}^{0}}{\sqrt{5}}+\frac{2p}{5\sqrt{3}}\,{}^{d}\mathcal{K}_{10}^{0}\right]
\\
{}^{d}\mathcal{K}_{32}^{0}(-\beta)
&=-\frac{5\sqrt{7}}{3 p}
\left[\frac{\gamma\,{}^{d}\mathcal{K}_{22}^{0}-\,{}^{d-1}\mathcal{K}_{22}^{0}}{\sqrt{5}}+\frac{2p}{5\sqrt{3}}\,{}^{d}\mathcal{K}_{12}^{0}\right],
}
\esub
which we obtained using the method described in Appendix~\ref{app:Kernel_raise}. To give an explicit example, we then have
\bealf{
\label{eq:S1_single_reorg}
\mathcal{\hat{S}}_{1}(\nu, p)&\approx 
-\oOx{}\, p-
\left[\frac{3}{2}\oOx{}+\frac{47}{25}\oOx{2}+\frac{7}{25}\oOx{3} \right]p^3
\\ \nonumber
&\!\!\!\!\!\!
+
\left[
\frac{5}{8}\oOx{}
-\frac{79}{50}\oOx{2}
-\frac{109}{50}\oOx{3}
-\frac{183}{350}\oOx{4} 
-\frac{11}{350}\oOx{5} 
\right]p^5
\\
\nonumber
\mathcal{\hat{S}}_{2}(\nu, p)&\approx 
\left[\frac{2}{3}\oOx{}+\frac{11}{30}\oOx{2} \right]p^2
+
\left[
\frac{7}{5}\oOx{2}
+\frac{6}{7}\oOx{3}
+\frac{19}{210}\oOx{4}
\right]p^4
}
for the dipole and quadrupole terms. 
To obtain the final momentum average one has to expand $f_{\ell}(\gamma, \gamma_{\rm p})/f(\gamma)$ in orders of $p$ and then collect terms as required.\footnote{The factor $f(\gamma)$ is isolated to ensure the convergence of the moments $\left<p^k\right>$ over the electron momenta.} With  Eq.~\eqref{app:MMrMB_k}, the results then follow trivially. To give a few examples, we find the pure velocity terms
\bealf{
\mathcal{\hat{S}}_{\rm kin}&\approx 
-\beta_{\rm p} \mu_{\rm p} \oOx{}
+
\frac{\beta_{\rm p}^2}{3}
\left\{
4\oOx{}+\oOx{2}
+
P_2(\mu_{\rm p})\left[2\oOx{}+\frac{11\oOx{2}}{10} \right]
 \right\}
\nonumber \\ \nonumber
&\!\!\!\!\!\!\!\!\!
-\beta_{\rm p}^3
\left\{
\mu_{\rm p}
\left[
2\oOx{}
+\frac{47\oOx{2}}{25}
+\frac{7\oOx{3}}{25}
\right]
+
P_3(\mu_{\rm p})\left[\frac{11\oOx{2}}{50}+\frac{13\oOx{3}}{150} \right]
\right\}
}
where $\mu_{\rm p}=\vbh_{\rm p}\cdot\vgh$ is the direction cosine of the peculiar velocity with respect to the photon direction.
The first term is the well-known (non-relativistic) kinematic SZ effect \citep{Sunyaev1980}, while those $\propto \beta_{\rm p}^2$ are leading order corrections from monopole and quadrupole scattering. The results agree with those of previous derivations noting that $\tau^*\equiv \tau/[1-\beta_{\rm p}\mu_{\rm p}]$, where $\tau$ is the optical depth as determined in the clusters rest frame \citep[see Sect.~4.4 of][for detailed explanation]{Chluba2012SZpack}. The third order terms have not been obtained before, but are given here for illustration. For typical cluster velocities these should be negligible.

To further compare with existing analytic expressions, we give the first order temperature correction to the kinematic SZ effect:
\bealf{
\label{eq:Dn_kinematic_T1}
\mathcal{\hat{S}}^{\The}_{\rm kin}
&\approx 
-\beta_{\rm p} \The \mu_{\rm p} 
\left[
10\oOx{}
+\frac{47\oOx{2}}{5}
+\frac{7\oOx{3}}{5}
\right].
}
This expression is consistent with Eq.~(28) of \citet{Chluba2012SZpack} once changing from $\tau^*$ to $\tau$ (giving rise to another term $\beta_{\rm p} \The \mu_{\rm p} Y_0$).  It also directly reproduces the term $C_1(x)$ of \citet{Nozawa1998SZ}. Similarly, we can confirm Eq.~(29) of \citet{Chluba2012SZpack} once transforming to $\tau$. We also directly reproduce the correction functions $C_2(x)$, $D_0(x)$ and $D_1(x)$ of \citet{Nozawa1998SZ}, noting that these use $\tau^*$ instead of the more meaningful rest frame optical depth \citep[see discussion in ][]{Chluba2012SZpack}.

\subsection{Kinematic corrections to third order in $\beta_{\rm p}$}
For illustration, we give all terms from octupole scattering up to $3^{\rm rd}$ order in the electron temperature:
\bsub
\bealf{
\label{eq:Octupole}
&\Delta n\big|^{\rm octupole}_{\rm SZ}\approx -\tau^* \beta_{\rm p}^3 P_3(\mu_{\rm p})\sum_{k=0}^5 \The^k O_k(x)
\\
&O_0=
\left[\frac{11\oOx{2}}{50}+\frac{13\oOx{3}}{150}\right]\,\nbb(x)
\\
&O_1=
\left[
\frac{96\oOx{2}}{25}
+\frac{151\oOx{3}}{25}
+\frac{2179\oOx{4}}{1050}
+\frac{89\oOx{5}}{525}
\right]\,\nbb(x)
\\
&O_2=
\Bigg[
\frac{744\oOx{2}}{25}
+\frac{5867\oOx{3}}{50}
+\frac{227303\oOx{4}}{2100}
+\frac{7229\oOx{5}}{210}
\nonumber \\
&\qquad \qquad 
+\frac{1443\oOx{6}}{350}
+\frac{83\oOx{7}}{525}
\Bigg]\,\nbb(x)
\\
&O_3=
\Bigg[
\frac{3204\oOx{2}}{25}
+\frac{241113\oOx{3}}{200}
+\frac{6603279\oOx{4}}{2800}
+\frac{65247\oOx{5}}{40}
\\ \nonumber 
&\qquad
+\frac{768716\oOx{6}}{1575}
+\frac{107048\oOx{7}}{1575}
+\frac{6743\oOx{8}}{1575}
+\frac{92\oOx{9}}{945}
\Bigg]\,\nbb(x).
}
\esub
These can be readily generated using the method described above. 
Higher order corrections can be obtained by extending the computation in a similar manner. This shows that the boost operator approach give an exact solution to the SZ effect.

\vspace{-3mm}
\section{Conclusions}
\label{sec:Conc}
In this work we demonstrated explicitly that the boost operator approach allows us to compute the SZ signals to all orders in the electron temperature and cluster's peculiar motion. The exact SZ operator, Eq.~\eqref{eq:kSZ_operator},
converts the collision integral into a differential operator in frequency,  generated by $\oOnu{}=-\nu\partial_\nu$. This can be easily used to compute the required signal functions to arbitrarily high order by using Taylor series expansions of the operators in orders of the electron momentum with replacements to compute the required momentum averaged, thereby reducing the computational burden significantly.
We confirmed our results using existing analytic expressions and also gave the new correction functions at third order of the cluster's peculiar motion.
Higher order corrections can be constructed using a {\tt Mathematica} notebook that can be found at \url{www.chluba.de/Mathematica}.

Although we have not explicitly considered the effect of observer motion on the SZ signal \citep{Chluba2005b, Nozawa2005,
Chluba2012SZpack}, the relevant transformation can be carried out by simple replacements of the variables \citep{Chluba2012SZpack}. We also mention that multiple scattering corrections \citep[e.g.,][]{Chluba2014mSZI, Chluba2014mSZII} as well as non-isothermal clusters can in principle be treated using the SZ operator, however, we do not expect any additional insights from a boost operator treatment. 

It may be instructive to derive the Kompaneets equation and its generalizations using the boost operator approach. For this, the scattering process in the rest frame has to include recoil effects and also Klein-Nishina corrections. Extensions to the scattering of anisotropic incoming radiation \citep[e.g.,][]{Chluba2012} as well as non-blackbody spectra \citep{HC2021, Lee2022RadSZ, Sabyr2022, Acharya2023CIBSZ} could also be of interest. 
Finally, the boost operator approach should also allow us to compute relativistic corrections to the polarized SZ effect \citep{Sunyaev1980, Sazonov1999, Itoh2000Pol} in an efficient way.
However, we leave these problems to future work.

\small 

\vspace{2mm}

\noindent
{\it Data Availability Statement}: {\tt Mathematica} files to reproduce some of the key results are available at \url{www.chluba.de/Mathematica}.

\bibliographystyle{mn2e}
\bibliography{Lit-2025}

\begin{thebibliography}{42}
\expandafter\ifx\csname natexlab\endcsname\relax\def\natexlab#1{#1}\fi

\bibitem[{{Acharya} \& {Chluba}(2023)}]{Acharya2023CIBSZ}
{Acharya} S.~K., {Chluba} J., 2023, \mnras, 519, 2138

\bibitem[{{ACT/DES/HSC Collaboration} {et~al}\mbox{.}(2025){ACT/DES/HSC
  Collaboration}, {Aguena}, {Aiola}, {Allam}, {Andrade-Oliveira}, {Bacon},
  {Bahcall}, {Battaglia}, {Battistelli}, {Bocquet}, {Bolliet}, {Bond},
  {Brooks}, {Calabrese}, {Carretero}, {Choi}, {da Costa}, {Costanzi},
  {Coulton}, {Davis}, {Desai}, {Devlin}, {Dicker}, {Doel}, {Duivenvoorden},
  {Dunkley}, {Ferraro}, {Flaugher}, {Frieman}, {Gallardo}, {Gatti},
  {Gaztanaga}, {Gill}, {Golec}, {Gruen}, {Gruendl}, {Halpern}, {Hasselfield},
  {Hill}, {Hilton}, {Hincks}, {Hinton}, {Hollowood}, {Honscheid}, {Hubmayr},
  {Huffenberger}, {Hughes}, {James}, {Klein}, {Knowles}, {Koopman}, {Kosowsky},
  {Lahav}, {Lee}, {Lin}, {Lokken}, {Madhavacheril}, {Plazas Malag{\'o}n},
  {Marrewijk}, {Marshall}, {McMahon}, {Mena-Fern{\'a}ndez}, {Miquel},
  {Miyatake}, {Mohr}, {Moodley}, {Mroczkowski}, {Naess}, {Nati}, {Nicola},
  {Niemack}, {Ogando}, {Oguri}, {Orlowski-Scherer}, {Page}, {Partridge}, {da
  Silva Pereira}, {Porredon}, {Qu}, {Ragavan}, {Ried Guachalla}, {Romer},
  {Carnero Rosell}, {Rykoff}, {Samuroff}, {Sanchez}, {Sevilla-Noarbe},
  {Sierra}, {Sif{\'o}n}, {Smith}, {Staggs}, {Suchyta}, {Swanson}, {Tucker},
  {Vargas}, {Vavagiakis}, {De Vicente}, {Weaverdyck}, {Weller}, {Wollack}, \&
  {Zubeldia}}]{ACTSZ2025}
{ACT/DES/HSC Collaboration} {et~al.}, 2025, arXiv e-prints, arXiv:2507.21459

\bibitem[{{Ade} {et~al}\mbox{.}(2019){Ade}, {Aguirre}, {Ahmed}, {Aiola}, {Ali},
  {Alonso}, {Alvarez}, {Arnold}, {Ashton}, {Austermann}, {Awan}, {Baccigalupi},
  {Baildon}, {Barron}, {Battaglia}, {Battye}, {Baxter}, {Bazarko}, {Beall},
  {Bean}, {Beck}, {Beckman}, {Beringue}, {Bianchini}, {Boada}, {Boettger},
  {Bond}, {Borrill}, {Brown}, {Bruno}, {Bryan}, {Calabrese}, {Calafut},
  {Calisse}, {Carron}, {Challinor}, {Chesmore}, {Chinone}, {Chluba}, {Cho},
  {Choi}, {Coppi}, {Cothard}, {Coughlin}, {Crichton}, {Crowley}, {Crowley},
  {Cukierman}, {D'Ewart}, {D{\"u}nner}, {de Haan}, {Devlin}, {Dicker},
  {Didier}, {Dobbs}, {Dober}, {Duell}, {Duff}, {Duivenvoorden}, {Dunkley},
  {Dusatko}, {Errard}, {Fabbian}, {Feeney}, {Ferraro}, {Flux{\`a}}, {Freese},
  {Frisch}, {Frolov}, {Fuller}, {Fuzia}, {Galitzki}, {Gallardo}, {Tomas Galvez
  Ghersi}, {Gao}, {Gawiser}, {Gerbino}, {Gluscevic}, {Goeckner-Wald}, {Golec},
  {Gordon}, {Gralla}, {Green}, {Grigorian}, {Groh}, {Groppi}, {Guan},
  {Gudmundsson}, {Han}, {Hargrave}, {Hasegawa}, {Hasselfield}, {Hattori},
  {Haynes}, {Hazumi}, {He}, {Healy}, {Henderson}, {Hervias-Caimapo}, {Hill},
  {Hill}, {Hilton}, {Hilton}, {Hincks}, {Hinshaw}, {Hlo{\v{z}}ek}, {Ho}, {Ho},
  {Howe}, {Huang}, {Hubmayr}, {Huffenberger}, {Hughes}, {Ijjas}, {Ikape},
  {Irwin}, {Jaffe}, {Jain}, {Jeong}, {Kaneko}, {Karpel}, {Katayama}, {Keating},
  {Kernasovskiy}, {Keskitalo}, {Kisner}, {Kiuchi}, {Klein}, {Knowles},
  {Koopman}, {Kosowsky}, {Krachmalnicoff}, {Kuenstner}, {Kuo}, {Kusaka},
  {Lashner}, {Lee}, {Lee}, {Leon}, {Leung}, {Lewis}, {Li}, {Li}, {Limon},
  {Linder}, {Lopez-Caraballo}, {Louis}, {Lowry}, {Lungu}, {Madhavacheril},
  {Mak}, {Maldonado}, {Mani}, {Mates}, {Matsuda}, {Maurin}, {Mauskopf}, {May},
  {McCallum}, {McKenney}, {McMahon}, {Meerburg}, {Meyers}, {Miller},
  {Mirmelstein}, {Moodley}, {Munchmeyer}, {Munson}, {Naess}, {Nati},
  {Navaroli}, {Newburgh}, {Nguyen}, {Niemack}, {Nishino}, {Orlowski-Scherer},
  {Page}, {Partridge}, {Peloton}, {Perrotta}, {Piccirillo}, {Pisano},
  {Poletti}, {Puddu}, {Puglisi}, {Raum}, {Reichardt}, {Remazeilles},
  {Rephaeli}, {Riechers}, {Rojas}, {Roy}, {Sadeh}, {Sakurai}, {Salatino},
  {Sathyanarayana Rao}, {Schaan}, {Schmittfull}, {Sehgal}, {Seibert}, {Seljak},
  {Sherwin}, {Shimon}, {Sierra}, {Sievers}, {Sikhosana}, {Silva-Feaver},
  {Simon}, {Sinclair}, {Siritanasak}, {Smith}, {Smith}, {Spergel}, {Staggs},
  {Stein}, {Stevens}, {Stompor}, {Suzuki}, {Tajima}, {Takakura}, {Teply},
  {Thomas}, {Thorne}, {Thornton}, {Trac}, {Tsai}, {Tucker}, {Ullom},
  {Vagnozzi}, {van Engelen}, {Van Lanen}, {Van Winkle}, {Vavagiakis},
  {Verg{\`e}s}, {Vissers}, {Wagoner}, {Walker}, {Ward}, {Westbrook},
  {Whitehorn}, {Williams}, {Williams}, {Wollack}, {Xu}, {Yu}, {Yu}, {Zago},
  {Zhang}, {Zhu}, \& {Simons Observatory Collaboration}}]{SOWP2018}
{Ade} P. {et~al.}, 2019, \jcap, 2019, 056

\bibitem[{{Birkinshaw}(1999)}]{Birkinshaw1999}
{Birkinshaw} M., 1999, Phys.~Rep, 310, 97

\bibitem[{{Bleem} {et~al}\mbox{.}(2015){Bleem}, {Stalder}, {de Haan}, {Aird},
  {Allen}, {Applegate}, {Ashby}, {Bautz}, {Bayliss}, {Benson}, {Bocquet},
  {Brodwin}, {Carlstrom}, {Chang}, {Chiu}, {Cho}, {Clocchiatti}, {Crawford},
  {Crites}, {Desai}, {Dietrich}, {Dobbs}, {Foley}, {Forman}, {George},
  {Gladders}, {Gonzalez}, {Halverson}, {Hennig}, {Hoekstra}, {Holder},
  {Holzapfel}, {Hrubes}, {Jones}, {Keisler}, {Knox}, {Lee}, {Leitch}, {Liu},
  {Lueker}, {Luong-Van}, {Mantz}, {Marrone}, {McDonald}, {McMahon}, {Meyer},
  {Mocanu}, {Mohr}, {Murray}, {Padin}, {Pryke}, {Reichardt}, {Rest}, {Ruel},
  {Ruhl}, {Saliwanchik}, {Saro}, {Sayre}, {Schaffer}, {Schrabback},
  {Shirokoff}, {Song}, {Spieler}, {Stanford}, {Staniszewski}, {Stark}, {Story},
  {Stubbs}, {Vanderlinde}, {Vieira}, {Vikhlinin}, {Williamson}, {Zahn}, \&
  {Zenteno}}]{Bleem2015}
{Bleem} L.~E. {et~al.}, 2015, \apjs, 216, 27

\bibitem[{{Carlstrom} {et~al}\mbox{.}(2002){Carlstrom}, {Holder}, \&
  {Reese}}]{Carlstrom2002}
{Carlstrom} J.~E., {Holder} G.~P., {Reese} E.~D., 2002, \araa, 40, 643

\bibitem[{{Challinor} \& {Lasenby}(1998)}]{Challinor1998}
{Challinor} A., {Lasenby} A., 1998, \apj, 499, 1

\bibitem[{{Challinor} \& {Lasenby}(1999)}]{Challinor1999}
{Challinor} A., {Lasenby} A., 1999, \apj, 510, 930

\bibitem[{{Challinor} \& {van Leeuwen}(2002)}]{Challinor2002}
{Challinor} A., {van Leeuwen} F., 2002, \prd, 65, 103001

\bibitem[{{Chluba}(2011)}]{Chluba2011ab}
{Chluba} J., 2011, \mnras, 415, 3227

\bibitem[{{Chluba} \& {Dai}(2014)}]{Chluba2014mSZII}
{Chluba} J., {Dai} L., 2014, \mnras, 438, 1324

\bibitem[{{Chluba} {et~al}\mbox{.}(2014){Chluba}, {Dai}, \&
  {Kamionkowski}}]{Chluba2014mSZI}
{Chluba} J., {Dai} L., {Kamionkowski} M., 2014, \mnras, 437, 67

\bibitem[{{Chluba} {et~al}\mbox{.}(2005){Chluba}, {H{\"u}tsi}, \&
  {Sunyaev}}]{Chluba2005b}
{Chluba} J., {H{\"u}tsi} G., {Sunyaev} R.~A., 2005, \aap, 434, 811

\bibitem[{{Chluba} {et~al}\mbox{.}(2012{\natexlab{a}}){Chluba}, {Khatri}, \&
  {Sunyaev}}]{Chluba2012}
{Chluba} J., {Khatri} R., {Sunyaev} R.~A., 2012{\natexlab{a}}, \mnras, 425,
  1129

\bibitem[{Chluba {et~al}\mbox{.}(2023)Chluba, Kite, \&
  Ravenni}]{chluba_spectro-spatial_2023-I}
Chluba J., Kite T., Ravenni A., 2023, Journal of Cosmology and Astroparticle
  Physics, 2023, 026

\bibitem[{{Chluba} {et~al}\mbox{.}(2012{\natexlab{b}}){Chluba}, {Nagai},
  {Sazonov}, \& {Nelson}}]{Chluba2012SZpack}
{Chluba} J., {Nagai} D., {Sazonov} S., {Nelson} K., 2012{\natexlab{b}}, \mnras,
  426, 510

\bibitem[{{Chluba} \& {Ravenni}(2025)}]{ChlubaBO25}
{Chluba} J., {Ravenni} A., 2025, arXiv e-prints, arXiv:2505.02080

\bibitem[{{Chluba} {et~al}\mbox{.}(2013){Chluba}, {Switzer}, {Nelson}, \&
  {Nagai}}]{Chluba2012moments}
{Chluba} J., {Switzer} E., {Nelson} K., {Nagai} D., 2013, \mnras, 430, 3054

\bibitem[{{Dai} \& {Chluba}(2014)}]{Dai2014}
{Dai} L., {Chluba} J., 2014, \prd, 89, 123504

\bibitem[{{Holder} \& {Chluba}(2021)}]{HC2021}
{Holder} G., {Chluba} J., 2021, arXiv e-prints, arXiv:2110.08373

\bibitem[{{Itoh} {et~al}\mbox{.}(1998){Itoh}, {Kohyama}, \& {Nozawa}}]{Itoh98}
{Itoh} N., {Kohyama} Y., {Nozawa} S., 1998, \apj, 502, 7

\bibitem[{{Itoh} {et~al}\mbox{.}(2000){Itoh}, {Nozawa}, \&
  {Kohyama}}]{Itoh2000Pol}
{Itoh} N., {Nozawa} S., {Kohyama} Y., 2000, \apj, 533, 588

\bibitem[{{Jauch} \& {Rohrlich}(1976)}]{Jauch1976}
{Jauch} J.~M., {Rohrlich} F., 1976, {The theory of photons and electrons}.
  {Springer}

\bibitem[{{Lee} \& {Chluba}(2024)}]{Lee2024SZpack}
{Lee} E., {Chluba} J., 2024, \jcap, 2024, 040

\bibitem[{{Lee} {et~al}\mbox{.}(2022){Lee}, {Chluba}, \&
  {Holder}}]{Lee2022RadSZ}
{Lee} E., {Chluba} J., {Holder} G.~P., 2022, \mnras, 512, 5153

\bibitem[{{Mroczkowski} {et~al}\mbox{.}(2019){Mroczkowski}, {Nagai}, {Basu},
  {Chluba}, {Sayers}, {Adam}, {Churazov}, {Crites}, {Di Mascolo}, {Eckert},
  {Macias-Perez}, {Mayet}, {Perotto}, {Pointecouteau}, {Romero}, {Ruppin},
  {Scannapieco}, \& {ZuHone}}]{Mroczkowski2019}
{Mroczkowski} T. {et~al.}, 2019, \ssr, 215, 17

\bibitem[{{Nozawa} {et~al}\mbox{.}(1998){Nozawa}, {Itoh}, \&
  {Kohyama}}]{Nozawa1998SZ}
{Nozawa} S., {Itoh} N., {Kohyama} Y., 1998, \apj, 508, 17

\bibitem[{{Nozawa} {et~al}\mbox{.}(2005){Nozawa}, {Itoh}, \&
  {Kohyama}}]{Nozawa2005}
{Nozawa} S., {Itoh} N., {Kohyama} Y., 2005, \aap, 440, 39

\bibitem[{{Planck Collaboration} {et~al}\mbox{.}(2014){Planck Collaboration},
  {Ade}, {Aghanim}, {Armitage-Caplan}, {Arnaud}, {Ashdown}, {Atrio-Barandela},
  {Aumont}, {Baccigalupi}, {Banday}, \& et~al.}]{Planck2013SZ}
{Planck Collaboration} {et~al.}, 2014, \aap, 571, A20

\bibitem[{{Planck Collaboration} {et~al}\mbox{.}(2016){Planck Collaboration},
  {Ade}, {Aghanim}, {Arnaud}, {Ashdown}, {Aumont}, {Baccigalupi}, {Banday},
  {Barreiro}, {Bartlett}, {Bartolo}, {Battaner}, {Battye}, {Benabed},
  {Beno{\^\i}t}, {Benoit-L{\'e}vy}, {Bernard}, {Bersanelli}, {Bielewicz},
  {Bock}, {Bonaldi}, {Bonavera}, {Bond}, {Borrill}, {Bouchet}, {Bucher},
  {Burigana}, {Butler}, {Calabrese}, {Cardoso}, {Catalano}, {Challinor},
  {Chamballu}, {Chary}, {Chiang}, {Christensen}, {Church}, {Clements},
  {Colombi}, {Colombo}, {Combet}, {Comis}, {Couchot}, {Coulais}, {Crill},
  {Curto}, {Cuttaia}, {Danese}, {Davies}, {Davis}, {de Bernardis}, {de Rosa},
  {de Zotti}, {Delabrouille}, {D{\'e}sert}, {Diego}, {Dolag}, {Dole},
  {Donzelli}, {Dor{\'e}}, {Douspis}, {Ducout}, {Dupac}, {Efstathiou}, {Elsner},
  {En{\ss}lin}, {Eriksen}, {Falgarone}, {Fergusson}, {Finelli}, {Forni},
  {Frailis}, {Fraisse}, {Franceschi}, {Frejsel}, {Galeotta}, {Galli}, {Ganga},
  {Giard}, {Giraud-H{\'e}raud}, {Gjerl{\o}w}, {Gonz{\'a}lez-Nuevo},
  {G{\'o}rski}, {Gratton}, {Gregorio}, {Gruppuso}, {Gudmundsson}, {Hansen},
  {Hanson}, {Harrison}, {Henrot-Versill{\'e}}, {Hern{\'a}ndez-Monteagudo},
  {Herranz}, {Hildebrandt}, {Hivon}, {Hobson}, {Holmes}, {Hornstrup}, {Hovest},
  {Huffenberger}, {Hurier}, {Jaffe}, {Jaffe}, {Jones}, {Juvela},
  {Keih{\"a}nen}, {Keskitalo}, {Kisner}, {Kneissl}, {Knoche}, {Kunz},
  {Kurki-Suonio}, {Lagache}, {L{\"a}hteenm{\"a}ki}, {Lamarre}, {Lasenby},
  {Lattanzi}, {Lawrence}, {Leonardi}, {Lesgourgues}, {Levrier}, {Liguori},
  {Lilje}, {Linden-V{\o}rnle}, {L{\'o}pez-Caniego}, {Lubin},
  {Mac{\'\i}as-P{\'e}rez}, {Maggio}, {Maino}, {Mandolesi}, {Mangilli}, {Maris},
  {Martin}, {Mart{\'\i}nez-Gonz{\'a}lez}, {Masi}, {Matarrese}, {McGehee},
  {Meinhold}, {Melchiorri}, {Melin}, {Mendes}, {Mennella}, {Migliaccio},
  {Mitra}, {Miville-Desch{\^e}nes}, {Moneti}, {Montier}, {Morgante},
  {Mortlock}, {Moss}, {Munshi}, {Murphy}, {Naselsky}, {Nati}, {Natoli},
  {Netterfield}, {N{\o}rgaard-Nielsen}, {Noviello}, {Novikov}, {Novikov},
  {Oxborrow}, {Paci}, {Pagano}, {Pajot}, {Paoletti}, {Partridge}, {Pasian},
  {Patanchon}, {Pearson}, {Perdereau}, {Perotto}, {Perrotta}, {Pettorino},
  {Piacentini}, {Piat}, {Pierpaoli}, {Pietrobon}, {Plaszczynski},
  {Pointecouteau}, {Polenta}, {Popa}, {Pratt}, {Pr{\'e}zeau}, {Prunet},
  {Puget}, {Rachen}, {Rebolo}, {Reinecke}, {Remazeilles}, {Renault}, {Renzi},
  {Ristorcelli}, {Rocha}, {Roman}, {Rosset}, {Rossetti}, {Roudier},
  {Rubi{\~n}o-Mart{\'\i}n}, {Rusholme}, \& {Sandri}}]{PlanckSZ2016}
{Planck Collaboration} {et~al.}, 2016, \aap, 594, A24

\bibitem[{{Pointecouteau} {et~al}\mbox{.}(1998){Pointecouteau}, {Giard}, \&
  {Barret}}]{Pointecouteau1998}
{Pointecouteau} E., {Giard} M., {Barret} D., 1998, \aap, 336, 44

\bibitem[{{Rephaeli}(1995)}]{Rephaeli1995}
{Rephaeli} Y., 1995, \apj, 445, 33

\bibitem[{{Sabyr} {et~al}\mbox{.}(2022){Sabyr}, {Hill}, \&
  {Bolliet}}]{Sabyr2022}
{Sabyr} A., {Hill} J.~C., {Bolliet} B., 2022, \prd, 106, 023529

\bibitem[{{Sarkar} {et~al}\mbox{.}(2019){Sarkar}, {Chluba}, \&
  {Lee}}]{CSpack2019}
{Sarkar} A., {Chluba} J., {Lee} E., 2019, \mnras, 490, 3705

\bibitem[{{Sazonov} \& {Sunyaev}(1998)}]{Sazonov1998}
{Sazonov} S.~Y., {Sunyaev} R.~A., 1998, \apj, 508, 1

\bibitem[{{Sazonov} \& {Sunyaev}(1999)}]{Sazonov1999}
{Sazonov} S.~Y., {Sunyaev} R.~A., 1999, \mnras, 310, 765

\bibitem[{{Shimon} \& {Rephaeli}(2004)}]{Shimon2004}
{Shimon} M., {Rephaeli} Y., 2004, New Astronomy, 9, 69

\bibitem[{{Sunyaev} \& {Zeldovich}(1980)}]{Sunyaev1980}
{Sunyaev} R.~A., {Zeldovich} I.~B., 1980, \mnras, 190, 413

\bibitem[{{The Simons Observatory Collaboration} {et~al}\mbox{.}(2025){The
  Simons Observatory Collaboration}, {Abitbol}, {Abril-Cabezas}, {Adachi},
  {Ade}, {Adler}, {Agrawal}, {Aguirre}, {Ahmed}, {Aiola}, {Alford}, {Ali},
  {Alonso}, {Alvarez}, {An}, {Arnold}, {Ashton}, {Atkins}, {Austermann},
  {Azzoni}, {Baccigalupi}, {Baleato Lizancos}, {Barron}, {Barry}, {Bartlett},
  {Battaglia}, {Battye}, {Baxter}, {Bazarko}, {Beall}, {Bean}, {Beck},
  {Beckman}, {Begin}, {Beheshti}, {Beringue}, {Bhandarkar}, {Bhimani},
  {Bianchini}, {Biermann}, {Biquard}, {Bixler}, {Boada}, {Boettger}, {Bolliet},
  {Bond}, {Borrill}, {Borrow}, {Braithwaite}, {Brien}, {Brown}, {Bruno},
  {Bryan}, {Bustos}, {Cai}, {Calabrese}, {Calafut}, {Carl}, {Carones},
  {Carron}, {Challinor}, {Chanial}, {Chen}, {Cheung}, {Chiang}, {Chinone},
  {Chluba}, {Cho}, {Choi}, {Chu}, {Clancy}, {Clark}, {Clarke}, {Cleary},
  {Clements}, {Connors}, {Contaldi}, {Coppi}, {Corbett}, {Cothard}, {Coulton},
  {Crowley}, {Crowley}, {Cukierman}, {D'Ewart}, {Dachlythra}, {Datta},
  {Day-Weiss}, {de Haan}, {Devlin}, {Di Mascolo}, {Dicker}, {Dober}, {Doux},
  {Dow}, {Doyle}, {Duell}, {Duff}, {Duivenvoorden}, {Dunkley}, {Dutcher},
  {D{\"u}nner}, {Edenton}, {El Bouhargani}, {Errard}, {Fabbian}, {Fanfani},
  {Farren}, {Fergusson}, {Ferraro}, {Flauger}, {Foster}, {Freese}, {Frisch},
  {Frolov}, {Fuller}, {Galitzki}, {Gallardo}, {Galvez Ghersi}, {Ganga}, {Gao},
  {Garrido}, {Gawiser}, {Gerbino}, {Gerras}, {Giardiello}, {Gill}, {Gilles},
  {Giri}, {Gleave}, {Gluscevic}, {Goeckner-Wald}, {Golec}, {Gordon}, {Gralla},
  {Gratton}, {Green}, {Groh}, {Groppi}, {Guan}, {Gupta}, {Gu{\dh}mundsson},
  {Hagstotz}, {Hargrave}, {Haridas}, {Harrington}, {Harrison}, {Hasegawa},
  {Hasselfield}, {Haynes}, {Hazumi}, {He}, {Healy}, {Henderson}, {Hensley},
  {Hertig}, {Herv{\'\i}as-Caimapo}, {Higuchi}, {Hill}, {Hill}, {Hilton},
  {Hilton}, {Hincks}, {Hinshaw}, {Hlo{\v{z}}ek}, {Ho}, {Ho}, {Ho}, {Hoang},
  {Hoh}, {Hornecker}, {Hornsby}, {Hotinli}, {Huang}, {Huber}, {Hubmayr},
  {Huffenberger}, {Hughes}, {Idicherian Lonappan}, {Ikape}, {Irwin}, {Iuliano},
  {Jaffe}, {Jain}, {Jense}, {Jeong}, {Johnson}, {Johnson}, {Johnson}, {Jones},
  {Jost}, {Kaneko}, {Karpel}, {Kasai}, {Katayama}, {Keating}, {Keller},
  {Keskitalo}, {Kim}, \& {Kisner}}]{SimonsII2025}
{The Simons Observatory Collaboration} {et~al.}, 2025, arXiv e-prints,
  arXiv:2503.00636

\bibitem[{{Wright}(1979)}]{Wright1979}
{Wright} E.~L., 1979, \apj, 232, 348

\bibitem[{{Yasini} \& {Pierpaoli}(2017)}]{Yasini2017}
{Yasini} S., {Pierpaoli} E., 2017, \prd, 96, 103502

\bibitem[{{Zeldovich} \& {Sunyaev}(1969)}]{Zeldovich1969}
{Zeldovich} Y.~B., {Sunyaev} R.~A., 1969, \apss, 4, 301

\end{thebibliography}

\onecolumn

\begin{appendix}

\section{Thomson collision term}
\label{app:Thomson}
For resting electrons, the Compton scattering cross section is given by \citep[e.g.,][]{Jauch1976}
\begin{equation}
\label{eq:App_kerdef2}
\frac{\text{d}\sigma}{\text{d}\Omega} = \frac{3}{16\pi}\,\Bigg[\frac{\omega}{\omega_0}\Bigg]^2
\left[
1+\mu_{\rm{sc}}^2+
\left(\frac{\omega}{\omega_0} + \frac{\omega_0}{\omega}-2\right)\right],
\qquad
\frac{\omega}{\omega_0} = \frac{1}{1+\omega_0(1-\mu_{\rm{sc}})},
\end{equation}
where $\mu_{\rm{sc}}=\vgh_0\cdot\vgh$ is the direction cosine between the incoming and outgoing photon, with energies $\omega_0=h \nu_0/\me c^2$ and $\omega=h \nu/\me c^2$, respectively. Assuming that recoil can be omitted ($\omega_0\ll 1$ and $\omega/\omega_0\approx 1$), one then obtains the Thomson scattering cross section for resting electrons
\begin{equation}
\label{eq:s_Thomson}
\frac{\text{d}\sigma}{\text{d}\Omega} = \frac{3}{16\pi}\,
\left[
1+\mu_{\rm{sc}}^2\right]\equiv
\frac{1}{4\pi}\,
\left[1+\frac{1}{2}P_2(\mu_{\rm{sc}})\right],
\end{equation}
with Legendre polynomial $P_2(\mu_{\rm{sc}})=(3\mu_{\rm{sc}}^2-1)/2$.
The Thomson collision term in the electron rest frame is then given by
\begin{align}
\label{app:Boltz_dsigma}
\frac{\text{d}n(\omega_0, \vgh_0)}{\text{d}\tau}  
&= \int \text{d}\vgh\,\frac{\text{d}\sigma}{\text{d}\Omega}\,
\Big[n(\omega_0, \vgh)- n(\omega_0, \vgh_0)\Big],
\end{align}
where $\tau=\int c\Ne \sigT \id t$ is the rest frame Thomson optical depth. To obtain a coordinate-independent result, we write the cross section as
\begin{equation}
\label{eq:s_Thomson_coordinate_ind}
\frac{\text{d}\sigma}{\text{d}\Omega} \equiv
Y_{00}(\vgh_0)\,Y^*_{00}(\vgh)+\frac{1}{10}\sum_{m=-2}^2
Y_{2m}(\vgh_0)\,Y^*_{2m}(\vgh),
\end{equation}
using the addition theorem for spherical harmonics. Inserting this into Eq.~\eqref{app:Boltz_dsigma} and carrying out the integral over $\text{d}\vgh$ we then obtain
\begin{align}
\label{app:Boltz_dsigma_final}
\frac{\text{d}n(\omega_0, \vgh_0)}{\text{d}\tau}  
&= 
Y_{00}(\vgh_0)\,n_{00}(\omega_0)+\frac{1}{10}\sum_{m=-2}^2
Y_{2m}(\vgh_0)\,n_{2m}(\omega_0)
- n(\omega_0, \vgh_0)
=n_{0}(\omega_0, \vgh_0)+\frac{1}{10}n_{2}(\omega_0, \vgh_0)-n(\omega_0, \vgh_0),
\end{align}
where in the last step we introduced $n_\ell(\omega_0, \vgh_0)=\sum_{m=-\ell}^\ell Y_{\ell m}(\vgh_0)\,n_{\ell m}(\omega_0)$.

\section{Raising operations for kernel elements}
\label{app:Kernel_raise}
For our problem we require the kernel elements ${}^{d}\mathcal{K}_{\ell 0}^{0}(-\beta)$ and ${}^{d}\mathcal{K}_{\ell 2}^{0}(-\beta)$. Starting with 
\bealf{
\label{eq:Kernels_00}
{}^{d}\mathcal{K}_{00}^{0}(-\beta)
&= \int  \frac{1}{[\gamma(1+\beta \mu')]^d} \frac{\id\vghp}{4\pi}
=\frac{(\gamma+p)^{1-d}-(\gamma-p)^{1-d}}{2(1-d)p}
}
with $D=\gamma(1+\beta \mu')$ and thus $\mu'=(D-\gamma)/p$, using spherical harmonic relations we can write 
\bealf{
Y_{\ell, 0}(\vgh)&=
\frac{\mu'\, Y_{\ell-1, 0}(\vgh')}{C_\ell}-\frac{C_{\ell-1}\,Y_{\ell-2, 0}(\vgh')}{C_\ell}
=-\sqrt{\frac{4\ell^2-1}{\ell^2}}\,\frac{\gamma-D}{p}Y_{\ell-1,0}(\vgh')-\frac{\ell-1}{\ell}\sqrt{\frac{2\ell+1}{2\ell-3}}Y_{\ell-2,0}(\vgh')
}
with $C_\ell=\sqrt{\ell^2/(4\ell^2-1)}$. This then implies
\bealf{
\label{app:recursion}
^{d}\mathcal{K}_{\ell \ell'}^{0}(-\beta)
&= 
\int  \frac{Y_{\ell 0}(\vghp)\,Y_{\ell' 0}(\vgh)}{[\gamma(1+\beta \mu')]^d} \id\vghp
=
-\sqrt{\frac{4\ell^2-1}{\ell^2}}\,\frac{\gamma \,{}^{d}\mathcal{K}_{\ell-1, \ell'}^{0}(-\beta) -{}^{d-1}\mathcal{K}_{\ell-1, \ell'}^{0}(-\beta)}{p}-\frac{\ell-1}{\ell}\sqrt{\frac{2\ell+1}{2\ell-3}}
\,{}^{d}\mathcal{K}_{\ell-2, \ell'}^{0}(-\beta).
}
Starting from $^{d}\mathcal{K}_{00}^{0}(-\beta)$, this relation can be recursively used to compute all elements $^{d}\mathcal{K}_{\ell 0}^{0}(-\beta)$ and also $^{d}\mathcal{K}_{0 \ell'}^{0}(-\beta)\equiv {}^{2-d}\mathcal{K}_{\ell' 0}^{0}(\beta)$ due to symmetries of the aberration kernel \citep{Dai2014}. With the elements $^{d}\mathcal{K}_{0 \ell'}^{0}(-\beta)$ one can then use Eq.~\eqref{app:recursion} to obtain all remaining $^{d}\mathcal{K}_{\ell \ell'}^{0}(-\beta)$.

\end{appendix}

\end{document}